\documentclass[12pt]{article}

\usepackage[dvips]{graphicx}
\usepackage{amssymb}

\newcommand\pubnumber{XXX-2013}
\newcommand\pubdate{\today}

\def\beq {\begin{equation}}
\def\eeq {\end{equation}}
\def\bea {\begin{eqnarray}}
\def\eea {\end{eqnarray}}

\def\rar {\rightarrow}

\def\sb {\bar{s}}

\def\sb {\bar{s}}

\def\Kb {\bar{K}}

\def\Ob {\bar{\Omega}}

\def\dkpp {$D^+ \! \rar \! K^- \p^+ \p^+$ }

\def\sp {\!+\!}
\def\sm {\!-\!}

\def\cK {{\cal{K}}}

\def\cM {{\cal{M}}}

\def\p {\pi}

\def\Title#1{\begin{center} {\Large #1 } \end{center}}
\def\Author#1{\begin{center}{ \sc #1} \end{center}}
\def\Address#1{\begin{center}{ \it #1} \end{center}}

\newcommand\pubblock{\rightline{\begin{tabular}{l} \pubnumber\\
         \pubdate  \end{tabular}}}
\newenvironment{Abstract}{\begin{quotation}  }{\end{quotation}}
\newenvironment{Presented}{\begin{quotation} \begin{center} 
             PRESENTED AT\end{center}\bigskip 
      \begin{center}\begin{large}}{\end{large}\end{center} \end{quotation}}
\def\Acknowledgements{\bigskip  \bigskip \begin{center} \begin{large}
             \bf ACKNOWLEDGEMENTS \end{large}\end{center}}

\def\beq{\begin{equation}}
\def\eeq#1{\label{#1}\end{equation}}
\def\eeqn{\end{equation}}

\def\beqa{\begin{eqnarray}}
\def\eeqa#1{\label{#1}\end{eqnarray}}
\def\eeqan{\end{eqnarray}}

\let\bar=\overbar

\def\Dslash{\not{\hbox{\kern-4pt $D$}}}
\def\dslash{\not{\hbox{\kern-2pt $\del$}}}

\def\msb{{\bar{\ssstyle M \kern -1pt S}}}

\begin{document}
\begin{titlepage}
\pubblock

\vfill
\Title{\dkpp: heavy meson decays and final state interactions}
\vfill
\Author{ P. C. Magalh\~{a}es\footnote{patricia@if.usp.br}, \underline{M. R. Robilotta}}
\Address{ Instituto de F\'{\i}sica, Universidade de S\~{a}o Paulo,  
S\~{a}o Paulo, SP, Brazil, 05315-970;}
\Author{  K. S. F. F. Guimar\~{a}es, T. Frederico, W. S. de Paula}
\Address{ Instituto Tecnol\'ogico de Aeron\'autica, 
 S\~ao Jos\'e dos Campos, SP, Brazil, 12.228-900;}
 \Author{I. Bediaga, A. C. dos Reis}
 \Address{Centro Brasileiro de Pesquisas F\'{i}sicas,Rio de Janeiro, 
RJ, Brazil, 22290-180;}
 \Author{ C.M. Maekawa}
 \Address{ Instituto de Matem\'atica, 
 Estat\'{\i}stica e F\'{\i}sica, Universidade Federal 
 do Rio Grande, Rio Grande, RS, Brazil; Campus 
 Carreiros, PO Box 474, 96201-900;}  
\vfill
\begin{Abstract}
We show that final state  interactions are important in 
shaping Dalitz plots 
for the decay $D^+ \rar K^- \p^+ \p^+$.
The theoretical treatment of this reaction requires
a  blend of several weak and hadronic processes and 
hence it is necessarily involved. 
In this talk we present results from a calculation which is still 
in progress, but has already unveiled the role of important
dynamical mechanisms.
We do not consider explicit quark degrees of freedom and 
our study is performed within an effective hadronic framework.
In spite of the relatively wide window of energies available  
in the Dalitz plot for
the $D^+$ decay, we depart from $SU(3)\times SU(3)$ 
chiral perturbation theory and extend its range by means of
unitarization. 
Our present results, which concentrate on the vector weak vertex,
describe qualitative features of the modulus of the decay
amplitude and agrees well with its phase in the elastic region.
\end{Abstract}
\vfill
\begin{Presented}
 Flavor Physics and CP Violation (FPCP-2013),
Buzios, Rio de Janeiro, Brazil, May 19-24 2013.
\end{Presented}
\vfill
\end{titlepage}
\def\thefootnote{\fnsymbol{footnote}}
\setcounter{footnote}{0}
%

\section{motivation}

The $S$-wave $K^- \p^+$ sub-amplitude in the decay 
$D^+ \rar K^- \p^+ \p^+$, denoted by $[K^-\p^+]_{D^+}$
has been extracted from data by the E791\cite{E791k} 
and FOCUS\cite{FOCUS} collaborations.
A remarkable feature of the results is a  significant deviation 
between $[K^-\p^+]_{D^+}$ and elastic $K^- \p^+$ LASS data \cite{LASS}.
This clearly indicates that the dynamical relationship between
both types of processes is not simple and has motivated 
an effort by our group aimed at understanding the
origins of this problem. A schematic calculation was presented in \cite{BR} and here we concentrate on recent progress.

\section{basic ideas}

\begin{figure}[htb] 
\begin{center}
\includegraphics[width=0.6\columnwidth,angle=0]{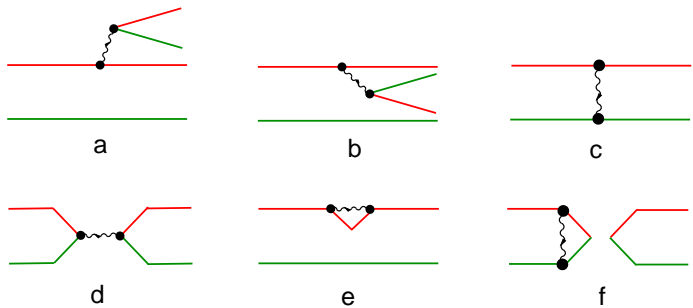}
\caption{Quark topologies for weak decays;
red lines are quarks, green lines are anti-quarks and
the wavy line is a $W\,$.}
\end{center}
\label{FChao}
\end{figure}

A suitable conceptual point of departure for dealing with weak
hadronic decays is the quark structures, 
described by Chao\cite{Chao}, based on 6 independent topologies
shown in Fig.\ref{FChao}.
The character of these diagrams is mostly symbolic,
since they do not include interactions mediated by gluons, 
but they implement properly the CKM quark mixing
for processes involving a single $W\,$.
Useful as they are, these diagrams provide just 
a bookkeeping for processes to be calculated in QCD
and in actual problems one needs to resort to 
form factors to be incorporated into hadronic effective lagrangians\cite{BSW}.

Calculations in non-perturbative QCD are  difficult 
and can only be performed by means of conceptual approximations. 
A rather powerful approximate scheme relies on the idea that 
the masses of the light quarks $u\,$, $d\,$, $s$ have just a limited 
relevance in processes involving low-energy mesons, so
that they can be treated perturbatively.
In the massless limit, the light sector becomes symmetric under
the $SU(3)\times SU(3)$ flavour group associated with chiral 
symmetry and the approximate scheme is known as chiral perturbation 
theory (ChPT) \cite{GL,EGPR}.
An important feature of this approach is that the light condensates
present in the vacuum are properly taken into account and 
pseudoscalar mesons are collective objects known as Goldstone 
bosons.
The programme is implemented by means of effective lagrangians,
which incorporate the symmetries of QCD whereas weak and electromagnetic 
interactions are included as external sources.
The inclusion of heavy mesons can be performed by suitable adaptations
of the light sector \cite{HM}.
At low energies, chiral perturbation theory provides the most precise
representation of QCD and unitarization of amplitudes,
supplemented by the use of form factors,
yields a reliable procedure for encompassing higher 
energies \cite{OO, Caprini}.

\section{dynamics}

\begin{figure}[htb]
\includegraphics[width=0.8\columnwidth,angle=0]{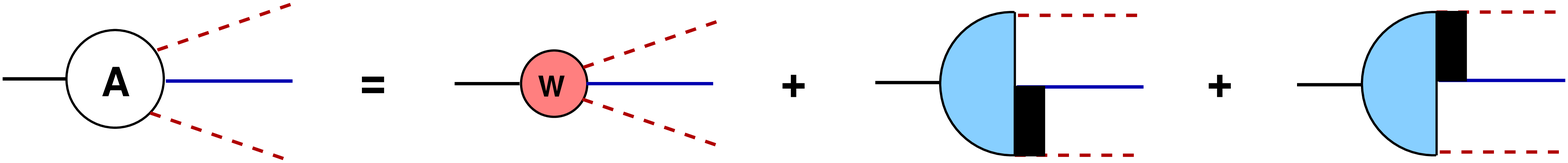}
\\[4mm]
\includegraphics[width=.96\columnwidth,angle=0]{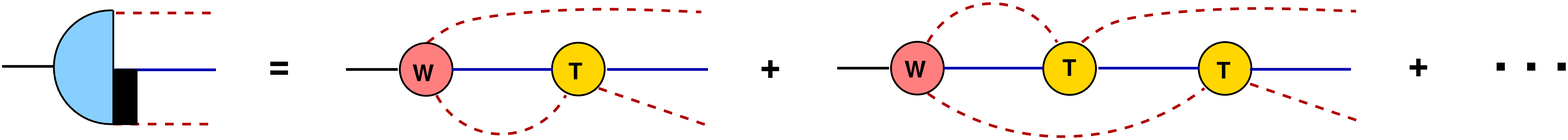}
\caption{Diagrammatic representation of the heavy meson decay
 into $K\pi\pi$, starting from the weak amplitude (red)
and including hadronic final state interactions.}
\label{FAmp}
\end{figure}

The reaction $D^+ \rar K^- \p^+ \p^+$ involves two distinct structures.
The first one concerns the primary quark transition $c\rar s \;W^+\,$,
which occurs in the presence of the light quark condensate 
of the QCD vacuum and is dressed into hadrons.
The second class of processes corresponds to three-body
final state interactions (FSIs), 
associated with the strong propagation of the state produced 
in the weak vertex to the detector, which eventually identifies a 
$K^- \p^+ \p^+$ state.
This amounts, in other words, to a kind of {\em hadronic interpolation}
between decay and detection.
These ideas are summarized in Fig.\ref{FAmp}.

\subsection{weak vertex}

\begin{figure}[htb] 
\begin{center}
\includegraphics[width=.8\columnwidth,angle=0]{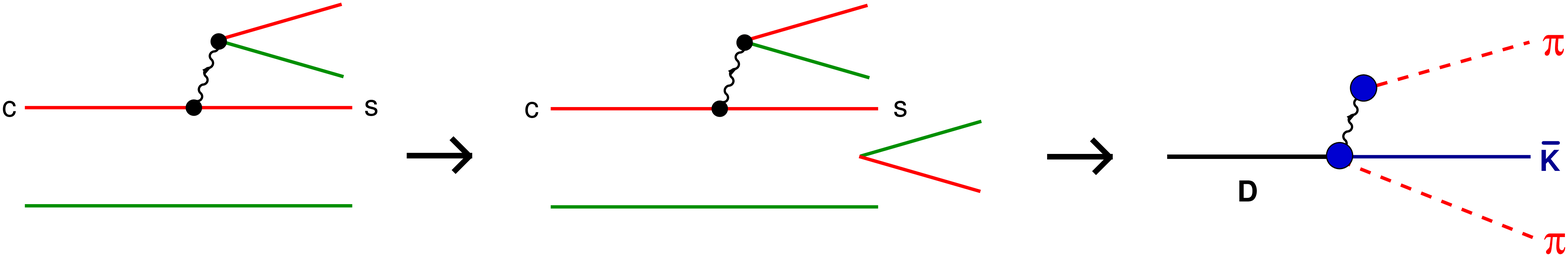}\\[10mm]
\includegraphics[width=.8\columnwidth,angle=0]{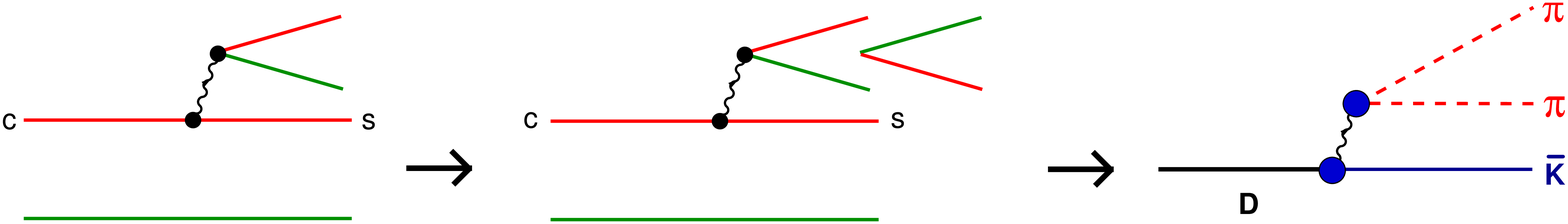}
\caption{Mechanisms for hadronization; quarks $u$ and $d$ are 
not specified.}
\end{center}
\label{FChaoH}
\end{figure}

The description of the weak vertex involves
just tree amplitudes and, 
in the case of the decay $D^+ \rar K^- \p^+ \p^+$, only diagrams $a$ and 
$b$ of Fig.\ref{FChao} contribute.
Here we concentrate just on the former class, which 
gives rise to the hadronic amplitudes shown in Fig.\ref{FChaoH}.
It is important to note that processes on the top involve
an {\em axial} weak current, whereas the bottom diagram is based on 
a {\em vector} current, since our results hint that the
latter dominates.
The blob in the diagrams summarizes several hadronic 
processes which contribute to form factors.

\begin{figure}[htb]
\begin{center}
\includegraphics[width=.9\columnwidth,angle=0]{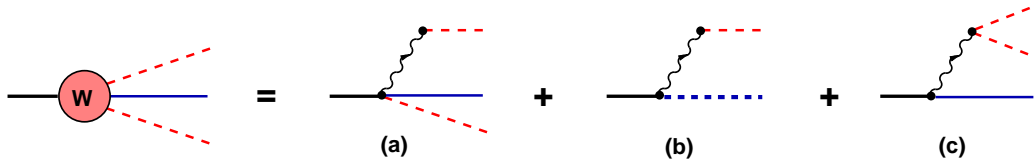}
\caption{Topologies for the weak vertex: the dotted line is a scalar 
resonance and the wavy line is a $W^+$, which is contracted to a point 
in calculations.}
\end{center}
\label{FAmpW}
\end{figure}

In the absence of form factors, the  weak vertex
entering Fig.\ref{FAmp} is given by the diagrams shown in Fig.\ref{FAmpW},
where processes $a$ and $b$ involve the axial current and $c$
contains a vector  current.
It is worth noting that one of the pions entering process $c$ is
neutral and hence this diagram does not contribute at tree level.

\begin{figure}[htb] 
\begin{center}
\includegraphics[width=1 \columnwidth,angle=0]{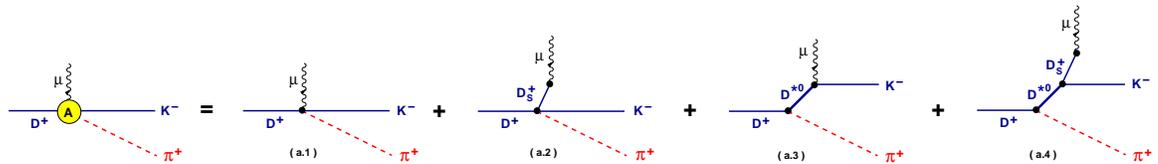}
\caption{Axial form factor.}
\end{center}
\label{FAFF}
\end{figure}

\begin{figure}[htb] 
\begin{center}
\includegraphics[width=.6\columnwidth,angle=0]{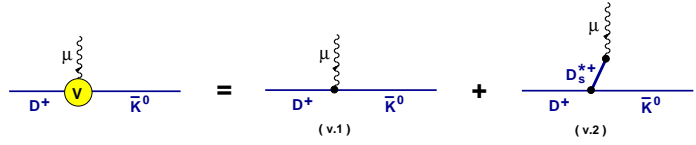}
\caption{Vector form factor.}
\end{center}
\label{FVFF}
\end{figure}

The inclusion of form factors can be made either 
by using phenomenological input or by  allowing the intermediate 
propagation of $(c\sb)$ states, as shown in Figs.\ref{FAFF} and 
\ref{FVFF}.

\newpage
\subsection{final state interactions}

\begin{figure}[htb]
\begin{center}
\includegraphics[width=0.7\columnwidth,angle=0]{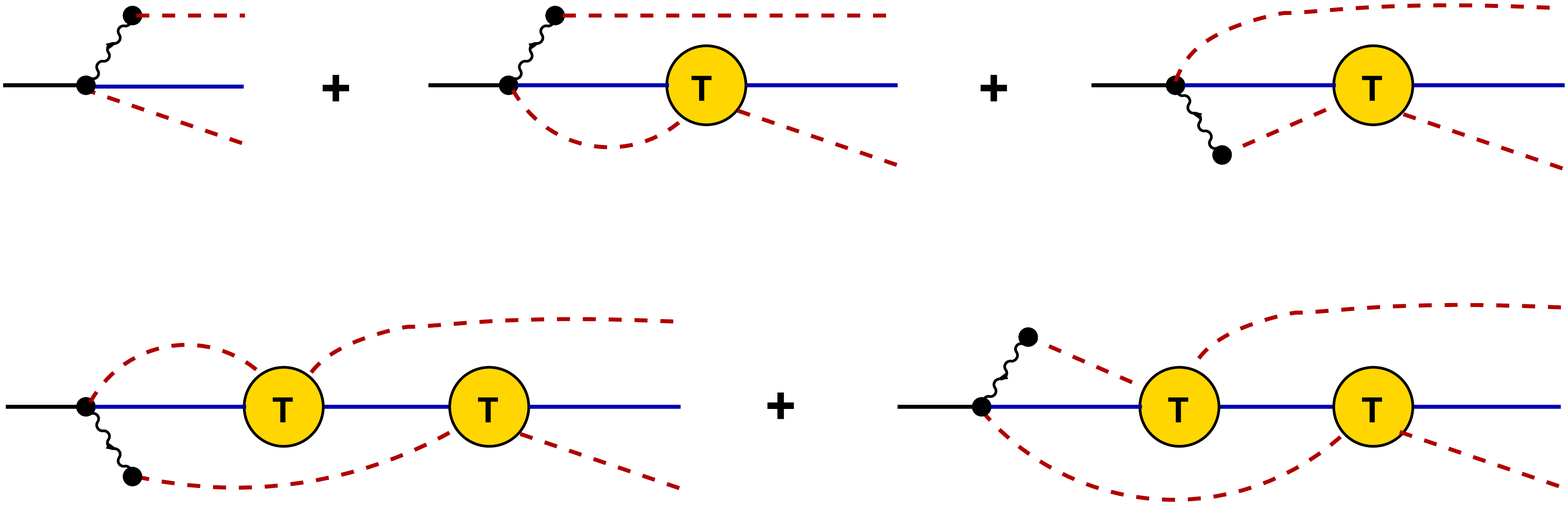}
\caption{Diagrams starting from the axial weak vertex $W_a$;
the wavy line is a $W^+$, always plugged to a $\p^+$;
the $\p$ produced together with the $\Kb$ in the opposite
side can be either positive or neutral.}
\end{center}
\label{FWA}
\end{figure}

\begin{figure}[htb]
\begin{center}
\includegraphics[width=0.8\columnwidth,angle=0]{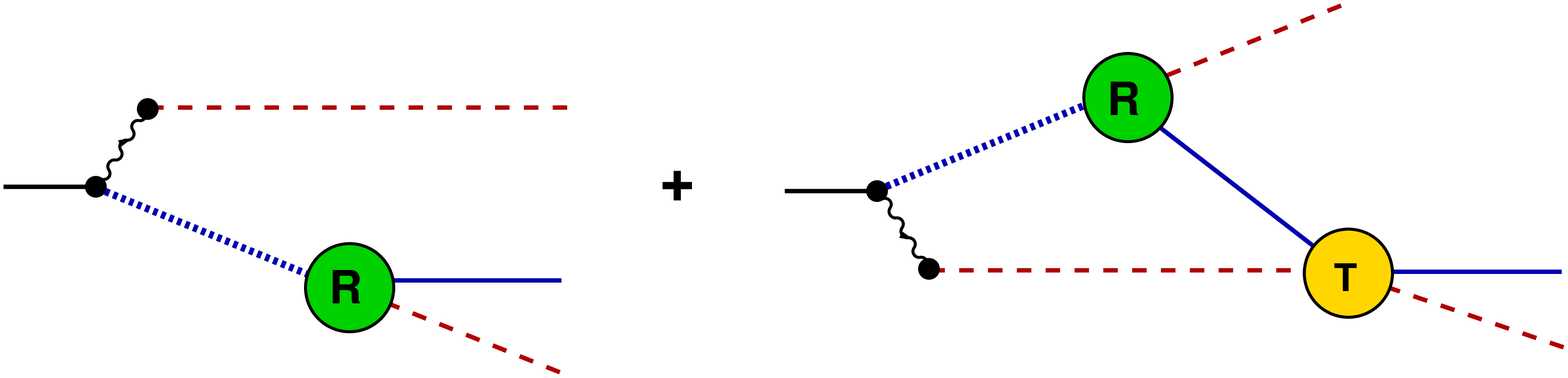}
\\[10mm]
\includegraphics[width=0.7\columnwidth,angle=0]{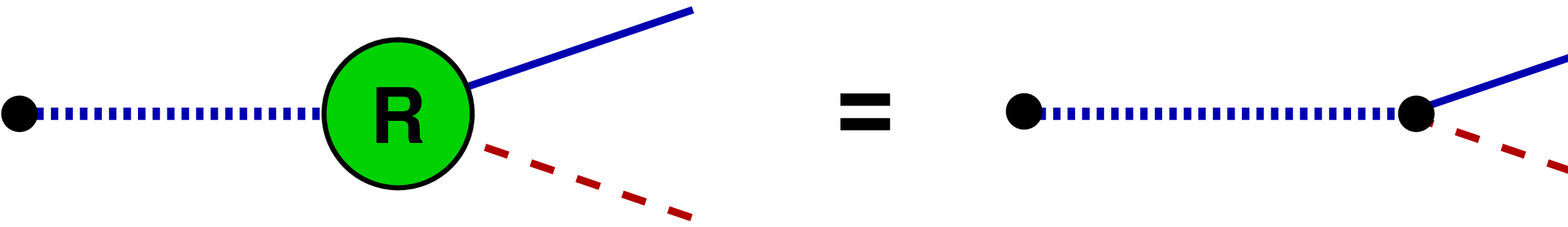}
\caption{Diagrams starting from the axial vertex $W_b$;
the wavy line is a $W^+$, always plugged to a $\p^+$
and the dotted line is a scalar resonance,
which has the width described at the bottom line.}
\end{center}
\label{FWB}

\vspace{10mm}

\begin{center}
\includegraphics[width=.8\columnwidth,angle=0]{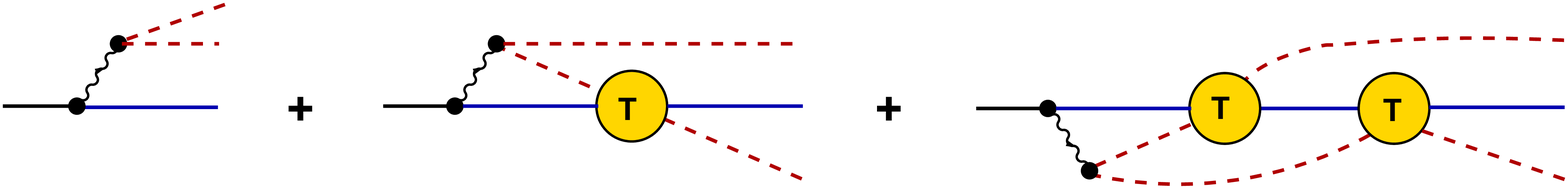}
\caption{Diagrams starting from the vector vertex $W_c$; 
one of the pions in the weak vertex is neutral.}
\end{center}
\label{FWC}
\end{figure}

When final state interactions are added  to the processes of 
Fig.\ref{FAmpW},
one finds three families of color-allowed
diagrams, as shown in Figs.\ref{FWA}-\ref{FWC}.
The $W^+$ is shown explicitly just for the sake of clarifying the 
various topologies, and is taken as a point in calculations.
The class of FSIs considered is based on a succession of elastic 
two-body interactions, which bring the $K \p$ phase into the problem.
Processes involving resonances have already been 
considered in Refs.\cite{DeD,Jap}
and quasi two-body axial FSIs have been discussed Ref.\cite{DiogoRafael}.
In our approach, the construction of the width of the
forward propagating resonance\cite{Diogo}
is displayed at the bottom of Fig.\ref{FWB}.
In the sequence, amplitudes corresponding to Figs.\ref{FWA}-\ref{FWC}
are denoted respectively by $A_a\,$, $A_b\,$, and $A_c\,$.

\newpage
\subsection{elastic $K \p$ amplitude}

The $K \p$ amplitude required in the construction of three-body FSIs 
is derived by means of chiral effective lagrangians,
based on leading order contact terms \cite{GL} and 
supplemented by resonances \cite{EGPR}, which allow for a wider 
energy range.
\begin{figure}[htb]
\begin{center}
\includegraphics[width=0.7\columnwidth,angle=0]{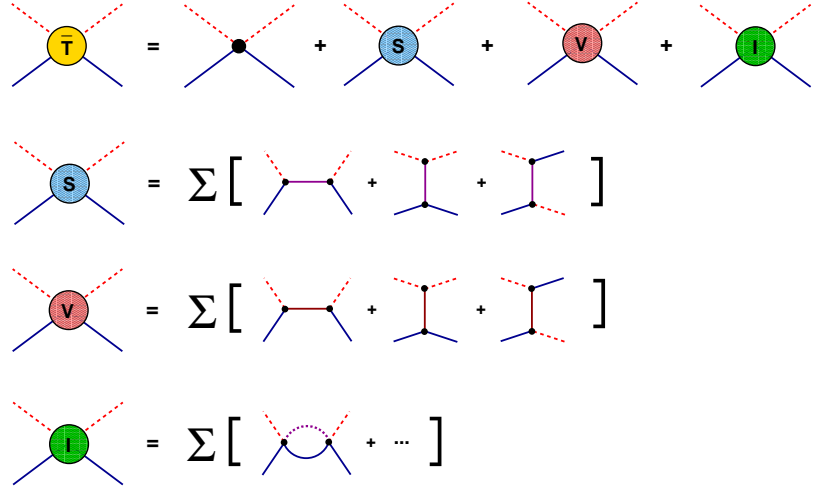}
\caption{ $K\p$ amplitude: the contact diagram is leading one at 
low-energies, 
whereas blobs are corrections due to other intermediate states; $S$ 
and $V$ correspond
to scalar and vector resonances and $I$, to inelastic channels; 
summation signs 
indicate the possibility of more than one intermediate state of each kind.}
\end{center}
\label{FKernel}
\end{figure}
Tree-level interactions, shown in Fig.\ref{FKernel},
give rise to a kernel $\cK\,$. The interaction of $\cK\,$  
in the Bethe-Salpeter equation\cite{OO} yields a 
unitary elastic amplitude $T\,$, given by 
\bea
T &\!=\!& \cK - \cK \,\Omega \,\cK 
+ \cK \,\Omega \,\cK \,\Omega \,\cK + \cdots
= \frac{\cK}{1+\Omega\,\cK}\;,
\label{e.1}
\eea 
where $\Omega$ is a two-meson propagator, which is 
a complex function above threshold.
Its real component is ultraviolet divergent and requires a subtraction.
The corresponding regular part is denoted by $\Ob\,$ and can 
be evaluated analytically.
After regularization, eq.(\ref{e.1}) becomes
\bea
T = \frac{\cK}{1+[C+\Ob]\,\cK}\;
\label{e.2}
\eea 
where $C$ is a constant adjusted to data.

If one wants, this result can be cast in a Breit-Wigner form, as
\bea
T = - \frac{16\p}{\rho} \; \frac{m\;\Gamma}
{(s- \cM^2) + i\,m\,\Gamma}\,,
\label{e.3}
\eea 
where $\rho=\sqrt{1\sm 2(M_\p^2 \sp M_K^2)/s \sp (M_\p^2 \sm M_K^2)^2}\;$,
$\cM$ is a running mass and $\Gamma$ is a width, which includes 
suitable kinematic factors.
The analytic extension of this amplitude to
the second Riemann sheet gives rise to poles.
If one does not include explicit resonances into the lagrangian,
one gets a single dynamically generated pole.
If just one explicit resonance is included, one gets a pair
of coupled poles, and so on.
This means that eq.(\ref{e.3}) always yields a dynamical pole,
which is rather broad and, in the $K \p$ system, is identified
as the $\kappa\;$.
This picture is consistent with chiral symmetry, since a low-mass
resonance cannot be accommodated into effective lagrangians.

\begin{figure}[htb]
\begin{center}
\includegraphics[width=0.5\columnwidth,angle=0]{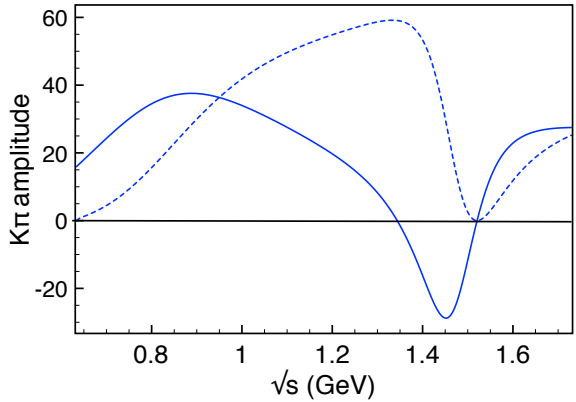}
\caption{Real (full line) and imaginary (dashed line) components of
the $K\p$ amplitude from eq.(\ref{e.2}).}
\end{center}
\label{FKPIamplitude}
\end{figure}

An important feature of the dynamical kernel produced by diagrams 
of Fig.\ref{FKernel} is that it vanishes for 
$\sqrt{s} \sim 1.5\,$GeV$\rar \,s \sim 2.3\,$GeV$^2$.
This zero propagates into the $K\p$ amplitude by means of eq.(\ref{e.2}),
as shown in Fig.\ref{FKPIamplitude} 
and therefore becomes a kind of signature of two-body
amplitudes in FSIs.

%

\section{first results}

In a previous publication \cite{BR}, we evaluated the contributions
of Figs.\ref{FWA}-\ref{FWC} to the $S$-wave $[K^-\p^+]_{D^+}$ 
sub-amplitude in the decay $D^+ \rar K^- \p^+ \p^+$.
With the purpose of taming the calculation, we made a number 
of simplifying assumptions.
Among them:
the weak amplitudes of Fig.\ref{FAmpW} were taken to be constants,
isospin $3/2$ and $P$ waves were not included in the $K\p$ amplitude
and couplings to either vector mesons or inelastic channels
were neglected.

\begin{figure}[htb]
\begin{center}
\includegraphics[width=0.6\columnwidth,angle=0]{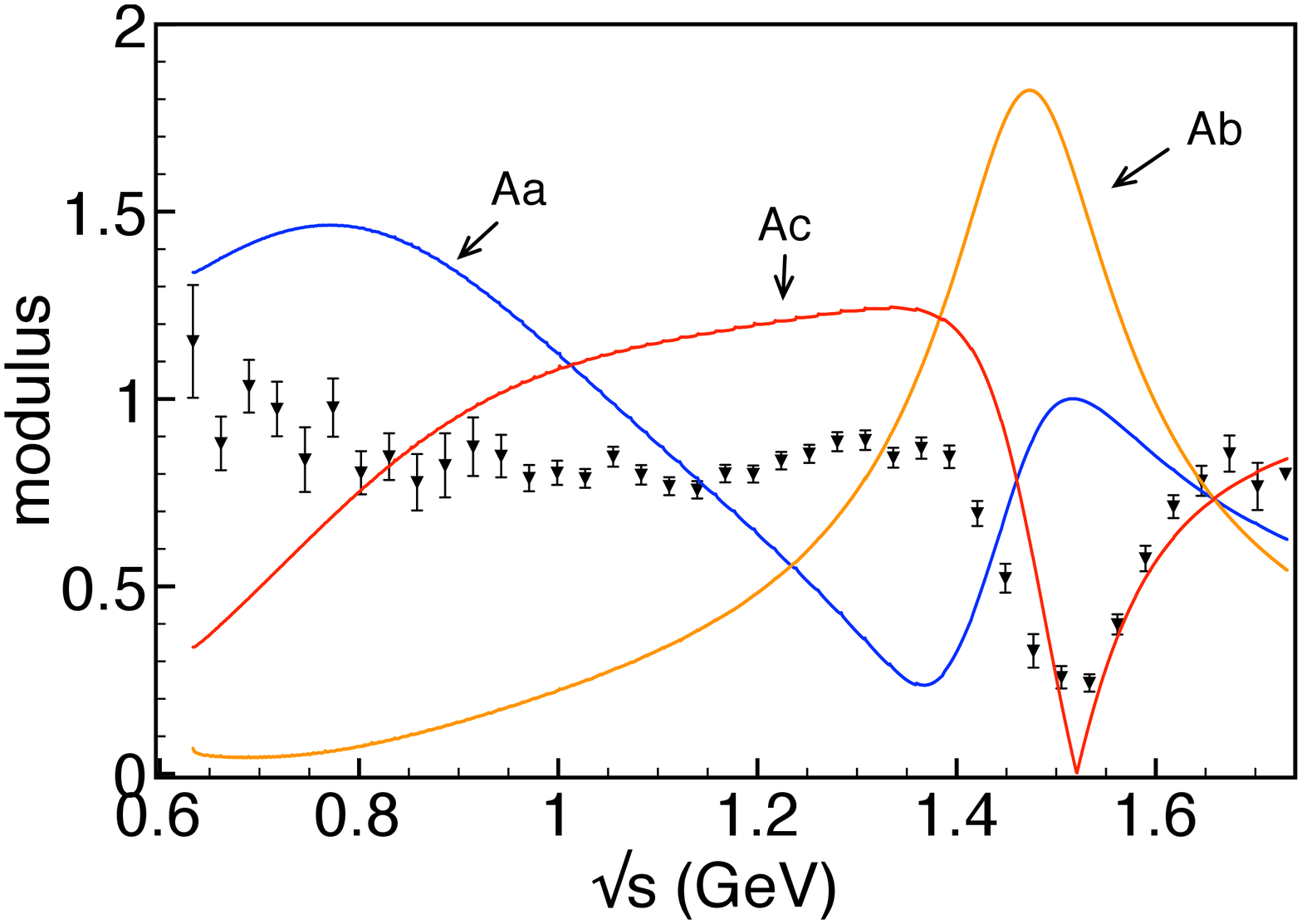}
\caption{ Behaviour of $|A_a|$, $|A_b|$ and $|A_c|$ compared with FOCUS data\cite{FOCUS}; scales are arbitrary.}
\end{center}
\label{Fmod}


%
\begin{center}
\includegraphics[width=0.85\columnwidth,angle=0]{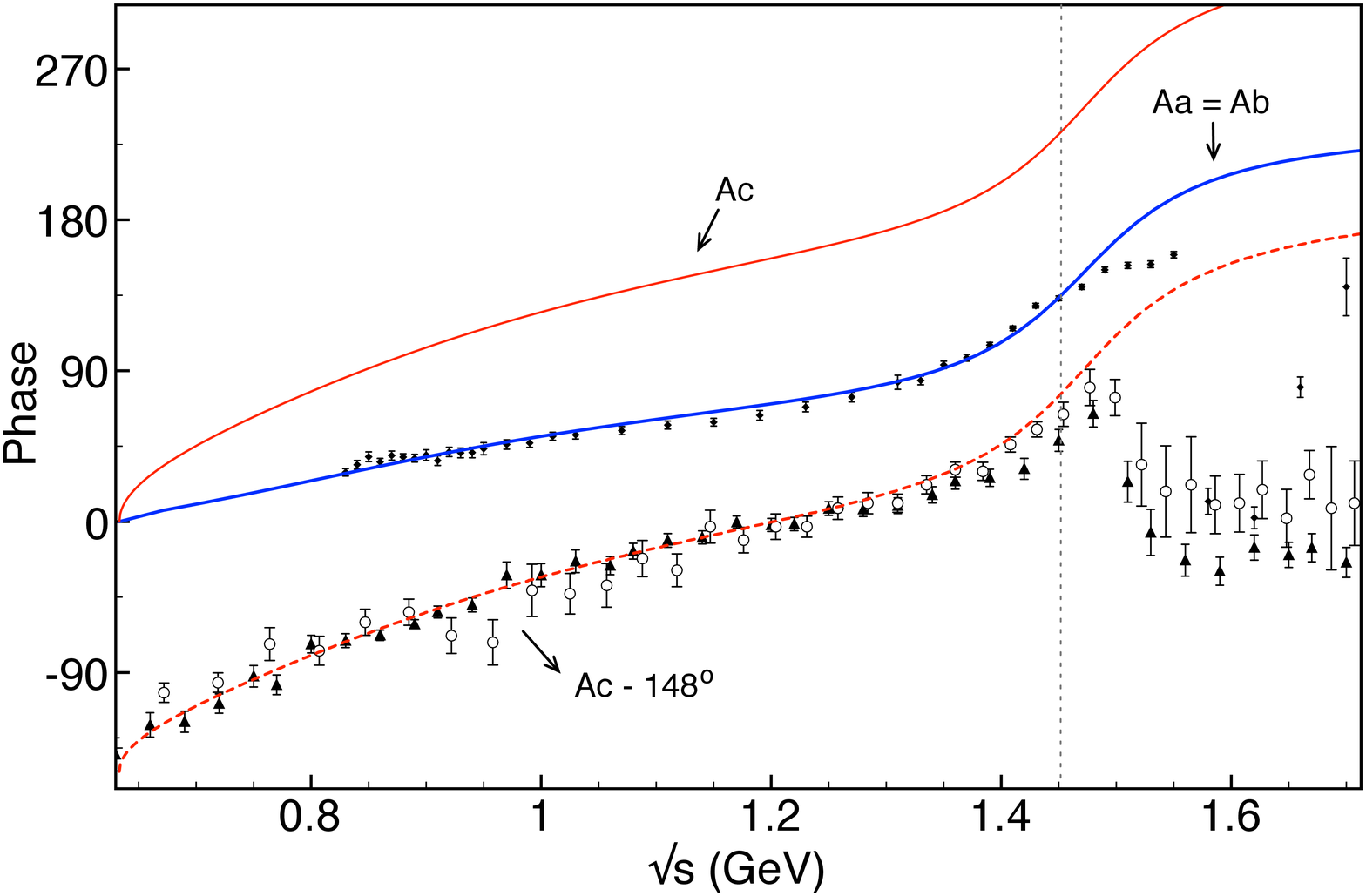}
\caption{Phase for $A_a$, $A_b$, $A_c$ at leading order and $A_c$ shifted by $-148^0$ compared with FOCUS 
\cite{FOCUS}(triangle) and E791\cite{E791k}(circle) data, together with elastic $K\pi$ results from LASS\cite{LASS}(diamond).}
\end{center}
\label{FShifted}
\end{figure}

Predictions for the modulus of $[K^-\p^+]_{D^+}$ are 
given in Fig.\ref{Fmod}.
Since all tree amplitudes in Fig.\ref{FAmpW} were assumed to be constants,
departures from horizontal straight lines in these figures are
signatures of final state interactions. As there is no vector contribution at tree level, one would have $A_c = 0$ in the absence of FSIs.
In the case of amplitudes $A_a$ and $A_c$, one finds dips
around $m_{12} \sim 1.5\,$GeV, whereas $A_b$ has a peak in 
that region.
Comparing these predictions with FOCUS data \cite{FOCUS},
one learns that the dip of the vector contribution $A_c$ 
occurs at the correct position, which is also the point
where the elastic $K \p$ amplitude vanishes, 
suggesting a direct correlation.
On the other hand, compatibility of data with $A_b\,$, which represents
the direct production of a 
resonance at the weak vertex, seems to be very difficult.

Results for the phase are displayed in fig.\ref{FShifted}.
Leading order contributions from the axial weak current,
represented by $A_a$ and $A_b\,$, obey Watson's theorem 
and fall on top of elastic $K\p$ data.
The curve for the vector $A_c$ amplitude
has a different shape and, if shifted by $-148^0$,
can describe well FOCUS data\cite{FOCUS}, up to the region of the peak.

The main lesson to be drawn from our first approach to this 
problem is that, for some yet unknown reason,
the amplitude which begins with a vector weak current, 
represented by diagram $(c)$ of Fig.\ref{FAmpW}, 
seems to be favoured by data.
This amplitude receives no contribution at tree level, since the
$W^+$ emitted by the charmed quark decays into a $\p^+\p^0$ pair.
Therefore, the leading term in this kind of process necessarily
involves loops and the corresponding imaginary components.
This motivates our present attempt to improve the description
of the vector vertex, discussed in the next section.

\section{vector vertex}

\begin{figure}[htb] 
\begin{center}
\includegraphics[width=0.5\columnwidth,angle=0]{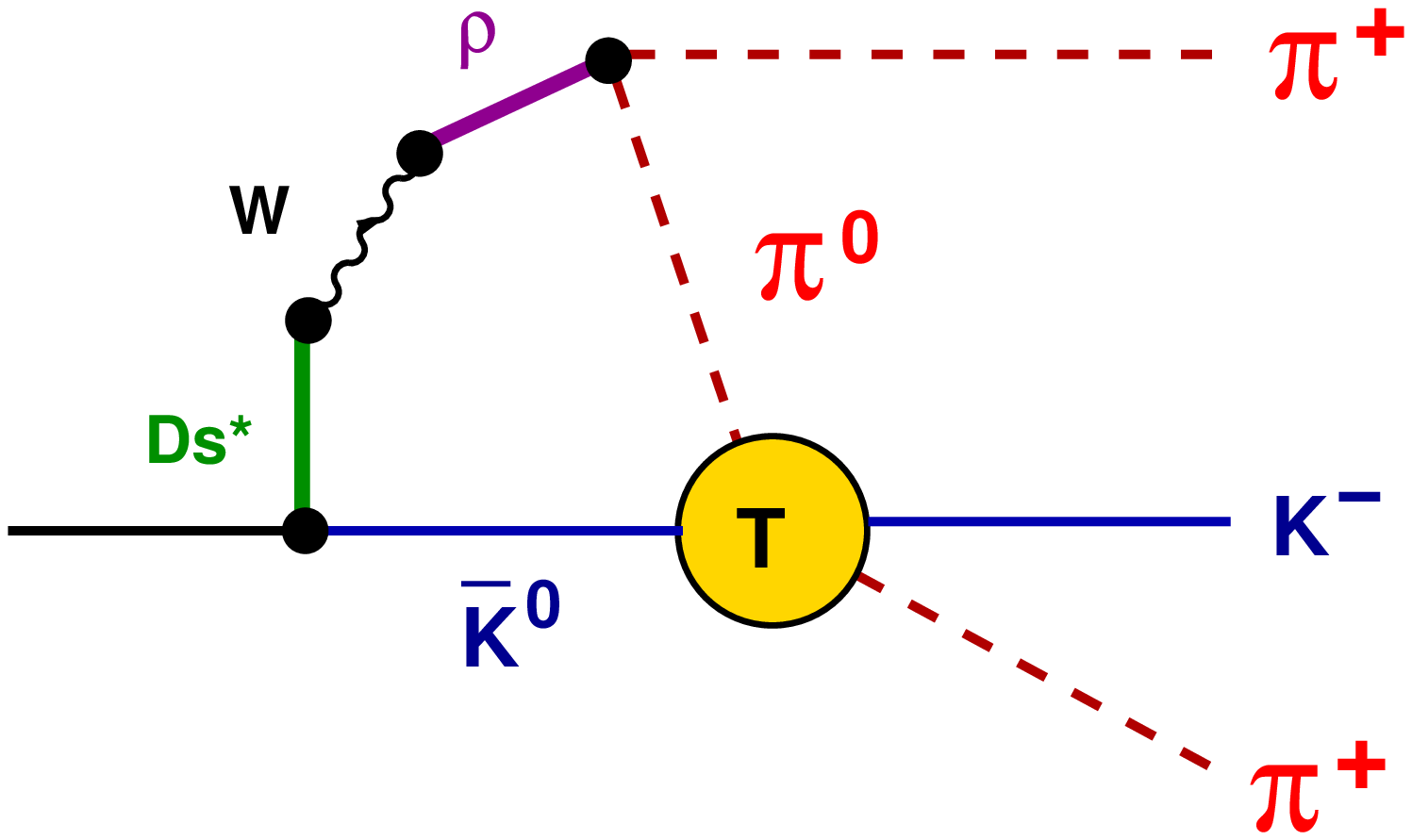}
\caption{Leading vector contribution.}
\end{center}
\label{Fvec}
\end{figure}

A limitation of our first study\cite{BR} of the reaction $D^+ \rar K^- \p^+ \p^+$
 was that all weak vertices 
were described by momentum independent functions.
Those results are now improved by considering:  the proper $P$-wave 
structure of the weak vertex, 
corrections associated with form factors and   
contributions from intermediate $\rho$ mesons.
Our basic interaction is described in Fig.\ref{Fvec}. 

%
The $\rho$ is introduced by means of standard vector meson dominance,
using the formalism given in Ref.\cite{EGPR}. 
The $D \rar W \Kb$ vertex may contain $(\sb c)$ intermediate states 
and can be obtained either by means of heavy-meson effective 
lagrangians\cite{HM} and the diagrams of 
Figs.\ref{FAFF}-\ref{FVFF}, or by using phenomenological information
parametrized in terms of nearest pole dominance \cite{weakFF}.
In the evaluation of fig.\ref{Fvec}, the $W$ is taken to be very heavy
and one is left with four hadronic propagators, which yield finite
values for loop integrals.

\begin{figure}[htb]
\begin{center}
\includegraphics[width=0.6\columnwidth,angle=0]{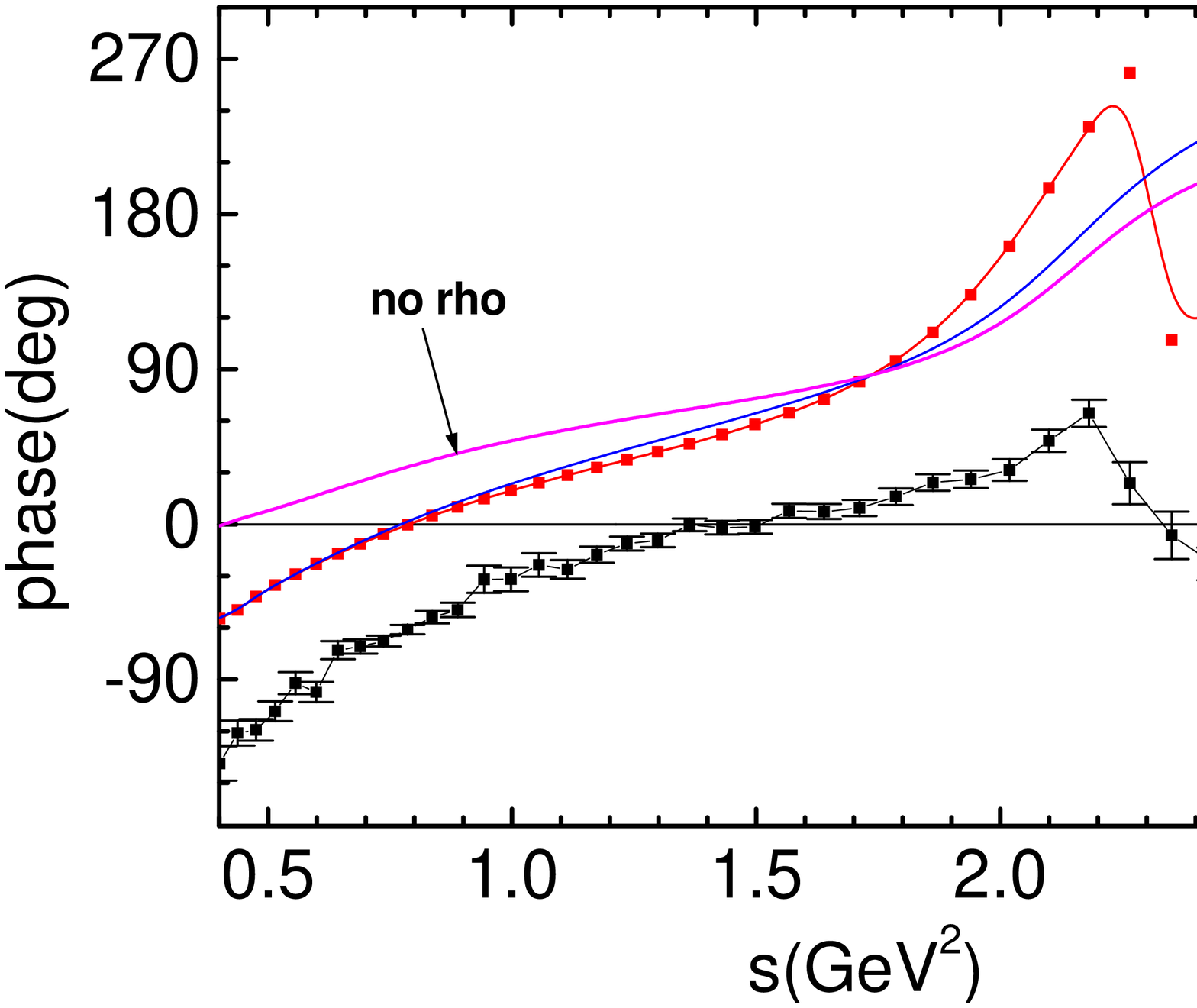}
\caption{Predictions for the phase, compared with FOCUS 
results\cite{FOCUS}.}
\end{center}
\label{FNPh}
\end{figure}

New predictions for the phase are shown in the red curve of Fig.\ref{FNPh}. 
Form factors, as expected, become more important at higher energies,
as indicated by the blue ``no form factors" curve.
The ``no rho" magenta curve is obtained by taking the limit 
$m_\rho \rar \infty$ in the calculation and tends to that labelled $A_c$ in Fig.\ref{FShifted}.
The most prominent feature of the full phase is that it
now has a negative value at threshold, showing that contributions
from light intermediate resonances are important. 
In QCD, loops are the only source of complex amplitudes. 
As the energy available in the loop of Fig.\ref{Fvec} may be larger
of both the $K\p$ and $K\rho$ thresholds, 
the amplitude has a  rich complex structure.
So far, the rho has just been treated as a point-like particle.
However its width, associated with two-pion intermediate states,
is a new source of complex amplitudes and is at present being included 
into the calculation.

\begin{figure}[htb]
\begin{center}
\includegraphics[width=0.6\columnwidth,angle=0]{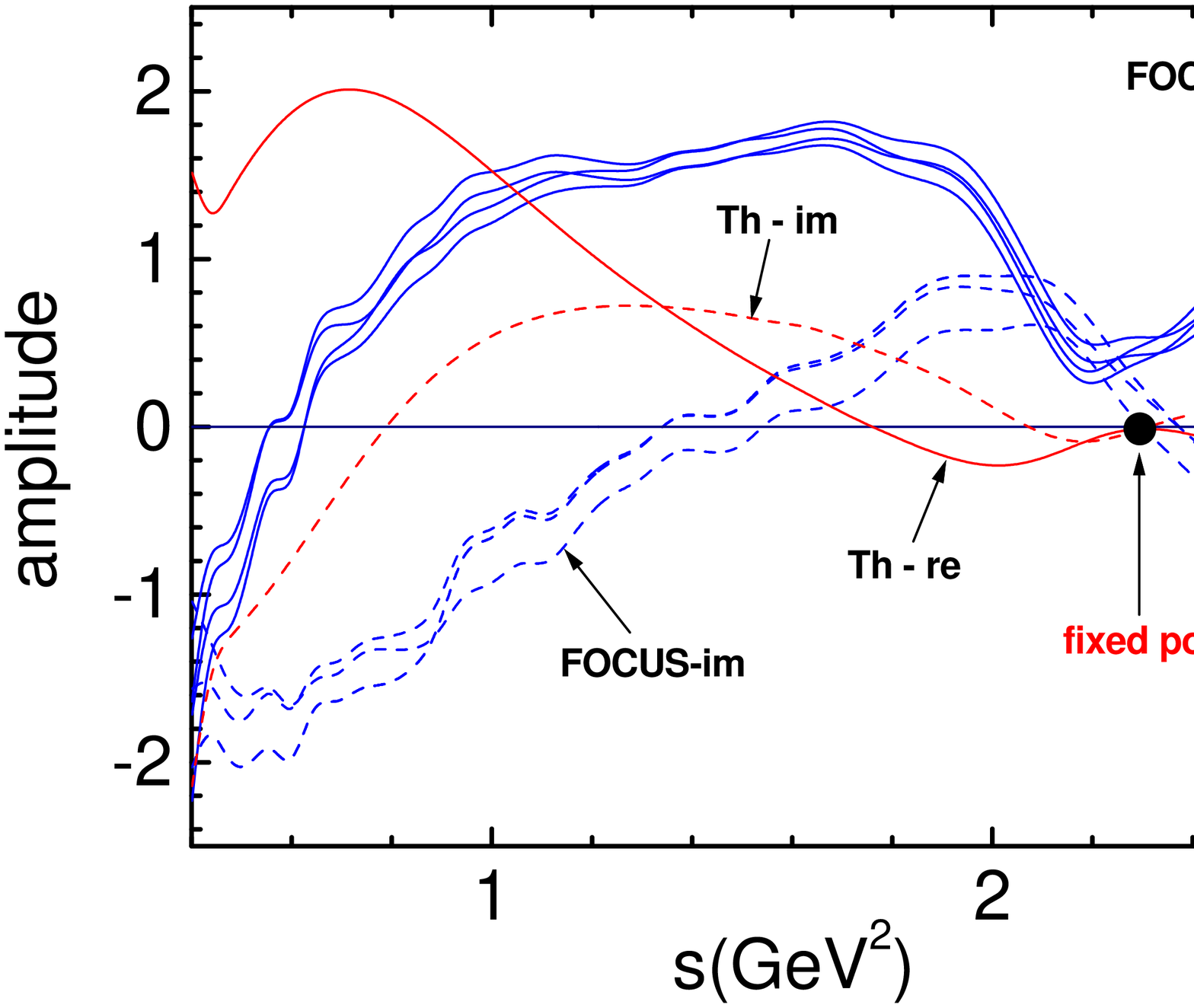}
\caption{Theoretical amplitude $[K^-\p^+]_{D^+}$ multiplied by 4 (red) compared with FOCUS \cite{FOCUS} data with error bars (blue).}
\end{center}
\label{FAmpFocus}
\end{figure}

In Fig.\ref{FAmpFocus} we display the real and imaginary components
of the $S$-wave $[K^-\p^+]_{D^+}$ amplitude determined by the 
FOCUS Collaboration \cite{FOCUS}.
A remarkable feature of these results is that both of them become 
very small around $s \sim 2.3\,$GeV$^2$.
This feature is related with the vanishing of the elastic $K\p$
amplitude, shown in Fig.\ref{FKPIamplitude}, which
occurs at the same energy. 
In fact, around that region and for $S$-waves, 
both figures suggest that $[K^-\p^+]_{D^+}$ is proportional 
to $ -i\,T_{K^-\p^+}\,$, equation (\ref{e.2}).
This seems to indicate the presence of loops and therefore
to confirm the dominance of weak vector currents in 
this branch of $D^+$ decays. 
For the sake of completeness, we also show the prediction
of the narrow rho to the $[K^-\p^+]_{D^+}$ amplitude, multiplied by 4, in order
to make it more visible.

\section{final remarks}

We have presented results for the decay $D^+ \rar K^- \p^+ \p^+$
and shown that final state  interactions are visible in data.
Even if our research programme is still incomplete and many aspects of the 
problem remain to be dealt with, it is already clear that
hadronic processes occurring between the primary weak decay 
and asymptotic propagation to the detector do play a key role in 
shaping experimental results.
Although derived from a single instance, the patterns of hadronic interpolation are quite general 
and it is fair to assume that
this conclusion can be extended to other processes. 
The treatment of this kind of problem 
involves many different aspects and is necessarily involved.
More theoretical effort on the subject would be definitely welcome \cite{Hanhart, meissberGardner, Bhattacharya, Doring}.

\Acknowledgements
MRR would like to thanks the  organizers of FPCP for the nice and productive conference. PCM was supported by FAPESP, process 09/50634-0.


\end{document}